\documentclass[12pt]{iopart}
\usepackage{graphicx}\usepackage{epsfig}\usepackage{color}
\usepackage{subfigure}\usepackage{lineno}
\usepackage{amsfonts}
\begin{document}
\title[Non-statistical decay and $\alpha$-correlations in the
$^{12}$C+$^{12}$C  ...]
{ Non-statistical decay and $\alpha$-correlations in the
$^{12}$C+$^{12}$C  fusion-evaporation reaction at 95 MeV }
%
%
\author{L Morelli$^{1}$, G Baiocco$^{1,2}$~\footnote{Present address:
Dipartimento di Fisica dell'Universit\`{a} and INFN, Pavia, Italy}, 
M D'Agostino$^{1}$,
F Gulminelli$^2$, \\ M Bruno$^{1}$, U Abbondanno$^3$, S Appannababu$^4$,
S Barlini$^{5,6}$, \\ M Bini$^{5,6}$, G Casini$^6$, M Cinausero$^4$, M Degerlier$^7$,\\
D Fabris$^8$, N Gelli$^6$, F Gramegna$^4$, V L Kravchuk$^{4,9}$,\\
T Marchi$^{4}$, A Olmi$^{6}$, G Pasquali$^{5,6}$, S Piantelli$^{6}$, 
S Valdr\'e$^{5,6}$ \\and Ad R Raduta$^{10}$ }
\address{$^1$Dipartimento di Fisica ed Astronomia dell'Universit\`{a} and INFN,
Bologna, Italy}
\address{$^2$LPC (IN2P3-CNRS/Ensicaen et Universit\'e),
F-14076 Caen c\'edex, France}
\address{$^3$INFN Trieste, Italy}
\address{$^4$INFN, Laboratori Nazionali di Legnaro, Italy}
\address{$^5$Dipartimento di Fisica ed Astronomia dell'Universit\`{a},
  Firenze, Italy}
\address{$^6$INFN Firenze, Italy}
\address{$^7$ University of Nevsehir, Science and Art Faculty, Physics
  Department, Nevsehir, Turkey}
\address{$^8$INFN, Padova, Italy}
\address{$^9$National Research Center \textquotedblleft Kurchatov
  Institute\textquotedblright, Moscow, Russia}
\address{$^{10}$NIPNE, Bucharest-Magurele, POB-MG6, Romania}
\ead{luca.morelli@bo.infn.it}
%
%
%
\begin{abstract}
Multiple alpha coincidences and correlations are studied in the
reaction $^{12}$C+$^{12}$C at 95 MeV for fusion-evaporation events
completely detected in charge. 
Two specific channels with Carbon and Oxygen residues in
coincidence with $\alpha-$particles are addressed, which are
associated with anomalously high branching ratios with respect to the
predictions by Hauser-Feshbach calculations.
Triple alpha emission appears kinematically compatible with a
sequential emission from a highly excited Mg. 
The phase space distribution of $\alpha-\alpha$ coincidences suggests a 
correlated emission from a Mg compound, leaving an Oxygen residue 
excited above the threshold for neutron decay.
These observations indicate a preferential $\alpha$ emission of 
$^{24}$Mg at excitation energies well above the threshold for $6-\alpha$ decay.
\end{abstract}
\noindent{\it NUCLEAR REACTIONS  12C(12C,X), E = 95 AMeV, 
Measured Fusion-evaporation reactions, Observed deviation from
statistical behaviour. Studied fragment-particle correlation
observables.\/} 

\submitto{\JPG}

\maketitle
\section{Introduction}
Since the first heuristic proposition of $\alpha$-chains as possible
building blocks of even-even nuclei in the late sixties~\cite{ikeda},
the subject of $\alpha$-clustering has been a central issue in
nuclear physics. It has even witnessed a gain of interest in recent
years~\cite{freer}. 
On the theoretical side, highly sophisticated ab-initio calculations
have shown pronounced cluster features in the ground state of a large
number of light nuclei~\cite{abinitio}. In addition, in
recent years, different constrained density functional approaches have
consistently found clear $\alpha$-cluster correlations in all light
and medium-heavy even-even nuclei at excitation energies around the
threshold of breakup into constituent
clusters~\cite{takemoto,maruhn,girod,vretenar}. 
Concerning experimental research, rotational bands consistent with
$\alpha$-cluster structures have been identified in different
even-even light nuclei and shown to persist even along their isotopic
chains~\cite{freer}. Exotic non-statistical decays of these correlated
states have been evidenced in the recent literature~\cite{cluster}. 

 A natural extension of the concept of nuclear clusters concerns
 nuclear molecules. Molecular states have been seeked for in nuclear
 reactions since the early days of heavy-ion science. In particular,
 several interesting resonances have been observed in the
 $^{12}$C+$^{12}$C reaction in the inelastic~\cite{inelastic} and
 $\alpha$-transfer channels~\cite{transfer}. These studies suggest
 that resonant structures persist in the $^{24}$Mg system up to
 around 50 MeV excitation energy. This is a surprising result as a pure
 statistical behaviour might be expected due to the extremely high
 number of available states at such high excitation. Concerning the
 $\alpha$-transfer channel, experimental results have been reproduced
 by coupled cluster calculations~\cite{takashina} where the cross
 section is dominated by a four-cluster
 ($\alpha+\alpha$)+($\alpha+^{12}$C) state of highly excited $^{24}$Mg
 around 30 MeV. Because of the remarkable persistence of cluster
 correlations at high excitation energies, the question naturally
 arises whether such correlations might affect other dissipative
 channels. These are typically associated with the formation of a
 compound nucleus, that is a system whose decay is assumed to be fully
 decoupled from the reaction entrance channel and governed by purely
 statistical laws. 

In a recent paper~\cite{prc} we have analyzed $^{12}$C+$^{12}$C
dissipative reactions at $95$ MeV and compared the experimental
data to the results of a dedicated Monte Carlo Hauser-Feshbach
code~\cite{baiocco} (HF$\ell$ from now on) 
for the evaporation of the CN $^{24}$Mg, at
$E^*/A=2.6$ MeV, issued in case of complete fusion. 
The angular momentum input distribution for the fused system in this
reaction is assumed to be a triangular one, with a maximum
value $J_{0\ max}=12\ \hbar$, coming from the systematics~\cite{PACE}.
Because of parity conservation, only even values of $J_0$ extracted
from the triangular distribution are allowed as an input for the CN
angular momentum.

We have shown that all the observables of dissipative events are fully
compatible with a standard statistical behaviour, with the exception
of $\alpha$-yields in coincidence with Oxygen residues. 

Specifically, the experimental Oxygen channel is dominated by the
presence of two $\alpha$ particles in coincidence, while the
Hauser-Feshbach theory predicts that the evaporation chains leading to
an Oxygen residue preferentially consist of one $\alpha$ and
two Hydrogen isotopes. 
A new data measurement of the same reaction has allowed us to analyze
these deviations in further details, and the results are reported in
the first paper~\cite{last_paper} of this series, hereby called paper I.
The new data set has confirmed the previous results, and
additionally shown an anomalously high branching ratio associated with
the $C$-$3\alpha$ channel. A cleaner event selection and a more
refined analysis led to an improved data reproduction by our
theoretical calculations.  
In the present work, which is a continuation of paper I, we analyse
the kinematical correlations in these specific channels to gather
further information on the emission mechanism. 

The paper is organized as follows. The observed deviations from a
standard statistical behaviour are summarized in \S\ref{deviations}. 
In the same section we show that such deviations
cannot be attributed to a memory of the entrance channel. 
\S\ref{correlations} deals with the multiple $\alpha$
correlations in non-statistical channels. 
Conclusions are drawn in \S\ref{conclusions}. 

\section{Deviations from a statistical behaviour}\label{deviations}
As explained in greater details in paper I, we have measured the
$^{12}$C+$^{12}$C reaction at 95 MeV with the 
GARFIELD+RCo set-up~\cite{epj}. We have selected a data-set consisting
of events completely detected in charge ($Z_{tot}=12$) which
corresponds to a large extent to the complete fusion-evaporation
channel. 
The statistical character of the data set is demonstrated by the good
reproduction of a very large set of inclusive and exclusive
observables by HF$\ell$ calculations.

Further details can be found in paper I, where we have shown
that the stringent condition $Z_{tot}=12$ does not artificially bias
the global event shape: the quality of the statistical model
reproduction of the different inclusive observables is the same with
$Z_{tot}\geq 10$ or $Z_{tot}=12$. 
However, an anomaly is observed in the probability of multiple $\alpha$
emission, as we show in the following. 

Figure~\ref{multa} summarizes the static observables related to
$\alpha$ production. First results were already shown
in~\cite{prc,last_paper}.
\subsection{Multiple $\alpha$ coincident yields}\label{multiplealpha}
\begin{figure}
\begin{center} 
\includegraphics[angle=0, width=0.75\columnwidth]{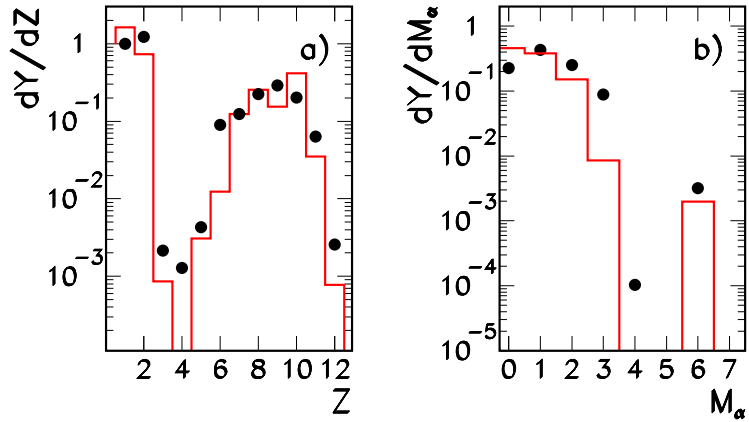}
\end{center}
\caption{(Color online) Left part: Inclusive charge distribution of
events completely detected in charge. For the same data set, the
right part displays the $\alpha$-particle multiplicity
distribution. Experimental data (symbols) are compared to filtered
HF$\ell$ calculations (lines).  
All distributions are normalized to the total number of events.
}
\label{multa}
\end{figure}
From the inclusive charge distribution displayed in the left part of
Figure~\ref{multa} we can see that the statistical model well
reproduces the global shape of the charge distribution, including the
$\alpha$ production. However, as shown in the right part of
Figure~\ref{multa}, it fails to reproduce the coincident yield ($M_{\alpha}$). 
Specifically, it underestimates the number of events with
$M_{\alpha}$\textgreater~1, while it overestimates those where no alphas are
emitted. 
The underestimation of 4 and 5 $\alpha$'s channels is clearly linked to the
missing yield of the lightest residues that we can see in the left
part of the figure. We have already discussed this feature in paper I
and tentatively attributed it to the lack of the fragment
break-up channel in the statistical model, channel which is expected
to be open at these high excitation energies (2.6 AMeV).  

At first sight, the presence of a peak in the multiplicity
distribution corresponding to the complete decomposition of the
system in constituent $\alpha-$particles ($M_{\alpha}=6$) could evoke a
vaporized phase~\cite{vaporization,texas} or a signature of Bose-Einstein
condensation~\cite{schuck}. However, the presence of such a peak is
predicted by a standard evaporation chain from an even-even compound,
due to the large branching ratio towards $\alpha$ decay at each step
of the de-excitation chain. 
This shows that the presence of a peak of
multiple-$\alpha$ coincidences is not a sufficient evidence of
$\alpha$ clustering, even if this possibility cannot be excluded. For
this reason, we will study the kinematical properties of these events
in greater detail in \S\ref{6a}. 

Another important effect observed in the inclusive charge distribution
is related to the strong underestimation of the Carbon production
yield. This could be due to a low predicted value for the $\alpha$ decay
probability of the parent nuclei (Ne, O, Mg), or alternatively to the
presence of direct reactions in the experimental sample. This reduced
Carbon yield can be associated with the underestimation of triple $\alpha$
coincidences observed in the right part of the figure, where there is
no gate on the residue charge. 
\begin{figure}[h]
\begin{center}
\includegraphics[angle=0, width=0.65\columnwidth]{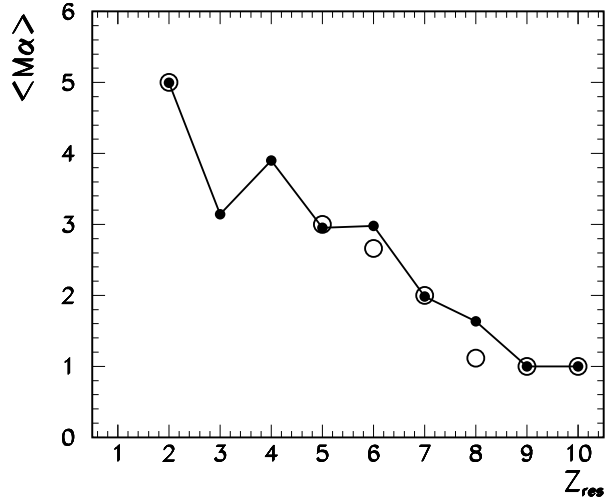}
\end{center}
\caption{Experimental (full circles) and predicted (open
circles) average $\alpha$-particle multiplicity as a function of the
charge of the heaviest fragment. 
Events where no $\alpha-$particles are emitted are excluded in this
analysis.} 
\label{mag}
\end{figure}

Moreover, looking at Figure~\ref{mag}, which displays
the average $\alpha$ yield associated with each residue~\footnote{No
theoretical point is associated with Z=3 and 4 because HF$\ell$ does not
produce a Z=3,4 as the highest Z residue of the evaporation chain},
it can be noticed that the $\alpha$ multiplicity associated with Carbon
residues is clearly higher in the data. This difference was already
shown in~\cite{last_paper}, where we have observed that the
measured branching ratio of the channel $C+3\alpha$
significantly overrates the predicted value. 
Figure~\ref{mag} shows another significative difference of branching
ratios between model and data concerning the Oxygen channel: the
$\alpha$ multiplicity in coincidence with Oxygen again overcomes the
statistical predictions. Similarly to the Carbon case, 
the dominant charge channel populated by HF$\ell$ in
coincidence with Oxygen is $(2H-\alpha-O)$, in contradiction with the
experimental measurement (see paper I). It is 
interesting to remark that this deviation is washed out in
the multiplicity distribution in Figure~\ref{multa}, because of the
large number of channels implying two coincident $\alpha-$particles. 

\begin{figure}[h]
\begin{center}
\includegraphics[angle=0, width=0.65\columnwidth]{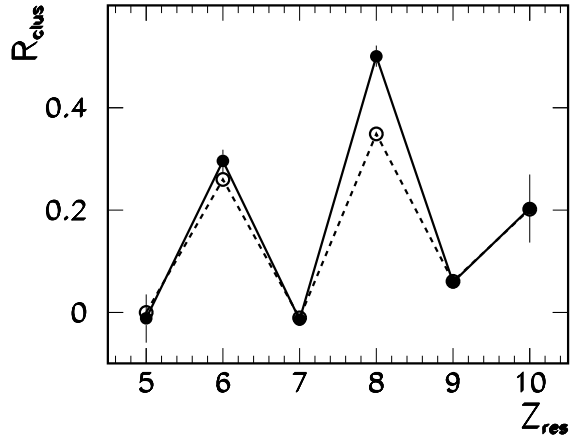}
\end{center}
\caption{Branching ratio excess for $\alpha$ decay
as a function of the atomic number of the final residue. Full symbols:
all events. Open symbols: more dissipative events only.
Lines are drawn to guide the eye.
All distributions are normalized to the unitary area. 
}
\label{deltag}
\end{figure}
To put in a better evidence possible $\alpha$ clustering
effects, we can define a variable quantifying the
experimental branching ratio excess for $\alpha$ emission:
\begin{equation}
R_{clus}(Z)=\frac{Y_{exp}(Z;n_Z\alpha)}{Y_{exp}(Z)}-\frac{Y_{HF\ell}(Z;n_Z\alpha)}{Y_{HF\ell}(Z)}
\label{delta}
\end{equation}
Here $Y(Z;n_Z\alpha)$ ($Y(Z)$) indicate coincident (inclusive) yields;
$n_Z\alpha$ is the (nearest integer) maximum $\alpha$ multiplicity
associated with the residue of charge $Z$ ($n_Z\alpha=(12-Z)/2$).
The subscripts \textquotedblleft{exp} and \textquotedblleft{HF$\ell$}
refer to experimental data and model predictions, respectively. 

The extra probability of $\alpha$ emission defined by (\ref{delta})
is plotted in Figure~\ref{deltag}. We can see that, in agreement with
Figure~\ref{mag}, the evaporation chains leading to a final Carbon or
Oxygen residue show a preferential $\alpha$ decay. A smaller effect in
the same direction is visible for a Neon residue.

A possible interpretation of this $\alpha$ excess may be due to the 
presence of residual $\alpha$ structure correlations in the excited
$^{24}$Mg or in its daughter nucleus $^{20}$Ne. We must notice that
the excitation energy of the compound $E^*(^{24}$Mg$)=62.4$ MeV is
well above the threshold for multiple-$\alpha$ decay, where such
correlations are expected. 
However the average $\alpha$ energy in the center of mass
results to be $10.5$ MeV and according to the HF$\ell$ calculation this
energy is increased to $13$ MeV for the first chance
emission. After the first 
evaporation step, considering the Q-value of the decay, the
daughter nucleus is expected to be at an excitation energy
$E^*(^{20}$Ne$)\approx 40$ MeV. This value is still much higher than
the energy threshold for 5-$\alpha$ dissociation 
$E^*_{theo}(^{20}$Ne$)\approx
16.74$ MeV where cluster states are theoretically
expected~\cite{maruhn,vretenar}, but cluster structures at energies
as high as 30 MeV have been already reported in the
literature~\cite{takashina}. 

To explore the possibility of alpha structure correlations in the
continuum, we now turn to study $\alpha$-channels 
in greater detail.

 \subsection{The influence of dissipation}
The well-known peculiarity of the $^{12}$C+$^{12}$C fusion-evaporation
reaction is that both the entrance channel and the compound nucleus
have potentially $\alpha$ structure correlations. In particular,
inelastic and $\alpha$ transfer reactions could be mixed to
fusion-evaporation events, thus explaining the failure of the
statistical model to correctly reproduce the Carbon and Oxygen
channel. 
\begin{figure}
\begin{center}
\includegraphics[angle=0, width=0.75\columnwidth]{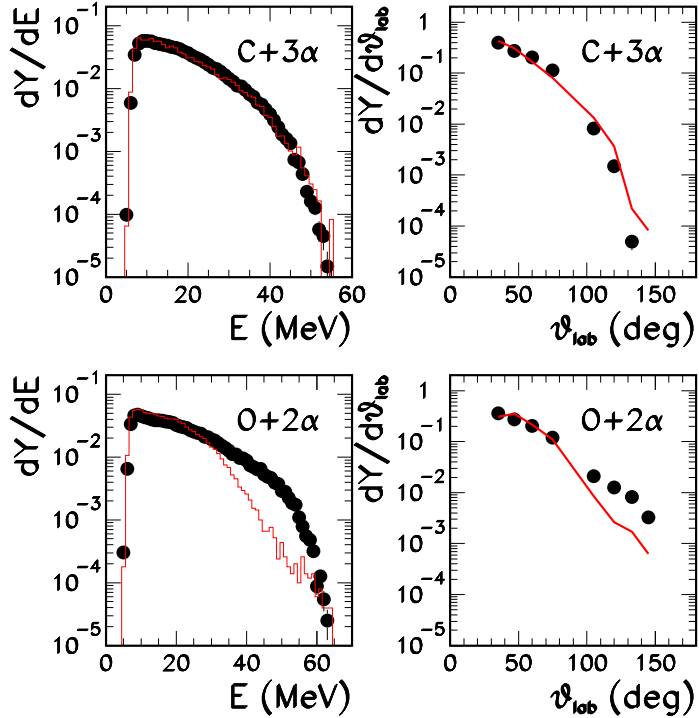}
\end{center}
\caption{(Color online)Energy spectra (left) and angular distributions (right) 
of $\alpha-$particles detected in coincidence with a Carbon (upper
part) or an Oxygen (lower part) residue. 
Data (symbols) are compared to HF$\ell$ calculations (lines). 
All distributions are normalized to the unitary area. 
}
\label{energy-angles}
\end{figure}
The distinction between direct and compound reactions is not
completely well defined. Indeed the physical processes are
continuous and the compound $^{24}$Mg could be in a
quasi-molecular state reminiscent of the entrance channel, even if it
is the source of $\alpha$ emission. 
A clear distinction between entrance channel effects and compound
effects can only be done by comparing the decay of the $^{24}$Mg
formed in two different entrance channels leading to the same
excitation energy.
To this aim, we have measured the reaction $^{14}$N$+^{10}$B at
$80.7$ MeV.
Altough the data are still under evaluation, preliminary
results~\cite{inpc} show indeed a reduction of the $(^A$O-2$\alpha)$
channel in this second reaction with respect to the $^{12}$C+$^{12}$C
one. A complete study is in preparation.

Direct reactions typically lead to angular distributions which are
reminiscent of  the entrance channel and therefore forward-backward
peaked in the laboratory system. This can be a hint to at least partially
disentangle the two mechanisms within a single data-set.
An indication in this sense comes from inspection of
Figure~\ref{energy-angles}. This figure shows the energy spectrum and
angular distribution of $\alpha-$particles emitted in coincidence
either with a Carbon or with an Oxygen residue, in comparison with the
HF$\ell$ calculation. 
While the Carbon channel does not show any indication of memory of the
entrance channel, a clear deviation is seen with respect to the
statistical model in the Oxygen case. Specifically, an excess of
backward emitted $\alpha-$particles (corresponding to low energy in the
laboratory frame) could indicate an incomplete memory loss of the
entrance channel, and a contamination from direct reactions 
in our experimental sample. 

To further explore this hypothesis, we study the $\alpha-$channels 
as a function of the dissipated energy. 
We introduce an estimate of the dissipation by the event-by-event quantity:
\begin{equation}
Q_{kin}=E_{kin}-E_{beam}=\sum_{i=1}^{N} E_i-E_{beam} \label{qdef}
\end{equation}
where $E_i$ is the energy of the particle in the laboratory frame and
the sum extends to the $N$ particles or fragments that are detected in 
coincidence, and exhausts the total charge $Z_{tot}=12$.
The quantity $Q_{kin}$ corresponds 
to the real $Q$-value of the reaction, given by the mass balance between 
the initial and final state, provided that the whole mass is collected in 
the outgoing channel (that is, in the absence of neutron emission). 

\begin{figure}
\begin{center}
\includegraphics[angle=0, width=1.1\columnwidth]{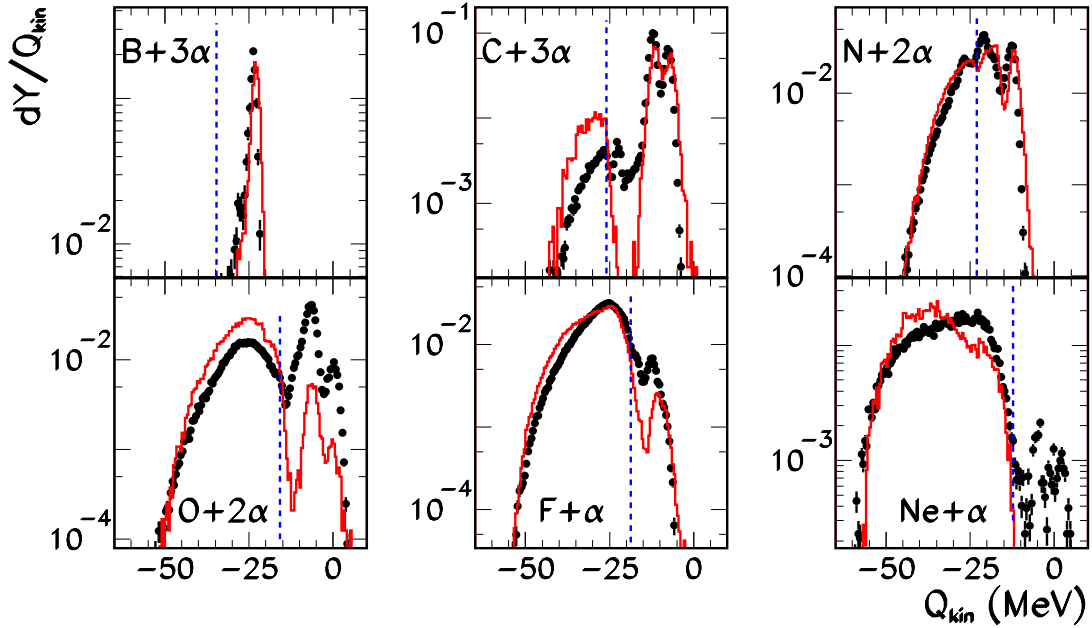}
\end{center}
\caption{(Color online) $Q_{kin}$ distributions in complete events
($Z_{tot}=12$) where the residue of charge from 5 to 10, indicated
in each panel, is detected in coincidence with the maximum possible
number of $\alpha-$particles.
Data (symbols) are compared to HF$\ell$ calculations (red lines). 
The dashed lines indicate the threshold Q-value for neutron emission
that we have adopted to separate more dissipative from
less dissipative events. All distributions are normalized to unitary area.
} 
\label{q_exp_12}
\end{figure}

Figure~\ref{q_exp_12} displays the experimental (symbols) and
theoretical (lines) $Q_{kin}$ distributions for the channels
corresponding to the maximum $\alpha$ multiplicity 
associated with the residue of charge $Z$. 

We can see that the theoretical and experimental 
spectra show a common structure, with narrow peaks at high $Q_{kin}$
values, and a broader region extending up to an important amount of
missing energy. 
In the model calculations, the different peaks correspond to the
various evaporation chains, starting from the $^{24}$Mg$^*$
compound nucleus, and finally leaving a residue of charge $Z$ in one
of its isotopic ground or low lying excited states. 
Despite the limited energy resolution in the $Q_{kin}$ reconstruction,
which broadens the peaks, the different levels can be clearly recognized in the
experimental sample and correspond to the predicted ones.

Starting from the threshold value of neutron emission associated with
each residue (dashed line in Figure~\ref{q_exp_12}), a continuous
contribution due to the missing neutron energy is superimposed to the
discrete $Q_{kin}$ spectrum. This contribution becomes dominant in
more dissipative events associated with low $Q_{kin}$ values, where neutron
emission from the continuum is the dominant decay mechanism.
We will call from now on, as in~\cite{last_paper}, $Q_<$ and $Q_>$ the
two regions below and above the neutron emission threshold,
respectively.

Besides the common pattern observed for the theoretical and experimental
$Q_{kin}$ distributions, clear differences are evident in the relative
population of the different regions, between experimental data and
calculations in Figure~\ref{q_exp_12}.
The larger deviations correspond to even-Z residues.
As in the case of the previous observables, this could be due to an
underestimation by the model of the $\alpha$ decay probability of even-Z
nuclei or due to the presence of direct reactions in the experimental sample. 

In particular, for the Neon residue, the model does not produce events
in the $Q_>$ region.
This deviation of the data from a statistical behaviour 
could be attributed to inelastic 
reactions due to quasi-molecular states of the compound $^{24}$Mg,
reminiscent of the entrance channel.
The comparison with another reaction~\cite{inpc}, where the same
compound nucleus is formed by non $\alpha$-like reaction partners
could shed some light on these discrepancies.

In the present paper we limit ourselves to the Oxygen and Carbon
residues, i.e. to (2$\alpha$-$^{A}$O$)$ and (3$\alpha$-$^{A}$C$)$ coincidences.
The highest $Q_{kin}$ peak ( $Q_{kin}\approx 0\ MeV$ for $O$,
$Q_{kin}\approx-7.3$ for $C$) corresponds to the respective ground state of
$^{16}$O and $^{12}$C. In the statistical calculation 
these peaks are obtained when the last-step $\alpha$ emission from the
$^{24}$Mg$^*$ compound nucleus leaves the residue directly in its ground 
state. 
The lower $Q_{kin}$ peaks ($Q_{kin}\approx-6.5\ MeV$ for $O$, 
$Q_{kin}\approx-11.7$ for $C$) correspond to the population of one of 
the particle bound excited states of $^{16}$O and $^{12}$C, 
which further $\gamma$-decay to the respective ground state. 

Starting from the threshold value for neutron emission
(${Q}_{kin}=-15.8\ MeV$ for $^{15}$O, ${Q}_{kin}=-26.0$ for
$^{11}$C) a broader distribution is observed due to neutron emission,
and the amount of missing energy corresponds to the undetected
neutron kinetic energy. 

Specifically, in the 2$\alpha$-$^{A}$O channel a much higher
percentage of events populates the $Q_{>}$ region in the experimental
sample with respect to model predictions. 
The larger experimental branching ratio for the multiple $\alpha$ 
exit channel in the $Q_{>}$ region is another indication of a 
possible contamination from direct reactions involving an excited
$^{12}$C nucleus, in competion with fusion-evaporation. 

\begin{figure}
\begin{center}
\includegraphics[angle=0, width=0.7\columnwidth]{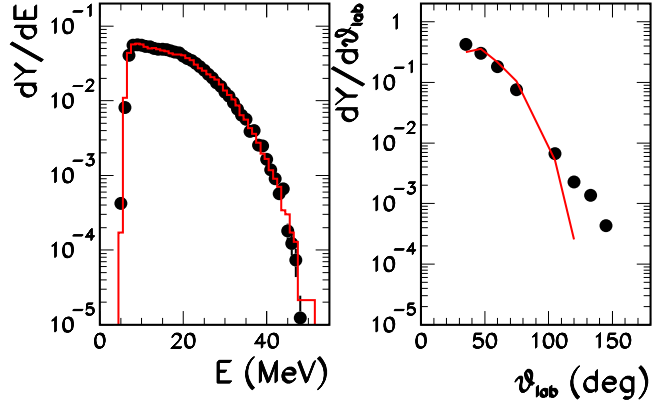}
\end{center}
\caption{(Color online) 
Energy spectra (left) and angular distributions (right) 
of $\alpha-$particles detected in coincidence with an Oxygen residue
in the 2$\alpha$-$^{A}$O events, for the most dissipative region $Q_<$
(see text). Data (symbols) are compared to HF$\ell$ calculations
(lines).  All distributions are normalized to the unitary area. 
}
\label{dissipative}
\end{figure}

Figure~\ref{dissipative} presents the $\alpha$ energy spectrum and the
angular distribution obtained when the less dissipative $(Q_>)$ events
are excluded from the analysis. 
We can see that the agreement between data and
model is improved, confirming the direct nature of the less
dissipative $Q_>$ events.

On the contrary in the 3$\alpha$-$^{A}$C channel the $Q_{kin}$
distribution is quite well reproduced by the model.
A slightly different proportion of more dissipative $Q_<$ versus 
less dissipative $Q_>$ can be observed between the two samples,
but the main discrepancy concerns the small peak between
$Q_{kin}=-20$ and $-24 \ MeV$, which is observed in the experimental
Carbon spectrum, and it is not visible in the calculation (see
Figure~\ref{q_exp_12} middle upper panel). 

In agreement with previous evaluations~\cite{prc}, this peak
corresponds to a contribution of about 3\% from 
the channel $^{13}$C$+^{3}$He+2$\alpha$ ($Q_{kin}=-22.9$ MeV).
This $^{3}$He$-\alpha$ contamination is due to $Z=2$ particles at the limit
of the mass identification threshold.

If we limit our attention to the true 3$\alpha$-$^{A}$C 
coincidences, we can conclude that the overall shape of the 
$Q_{kin}$ spectrum is consistent with a statistical behaviour.
The main significant deviation associated with Carbon is the overall higher 
experimental probability of the 3$\alpha$-$^{A}$C channel, independently
of the dissipation (see Figure~\ref{mag} and paper I).

\begin{figure}
\begin{center}
\includegraphics[angle=0, width=0.65\columnwidth]{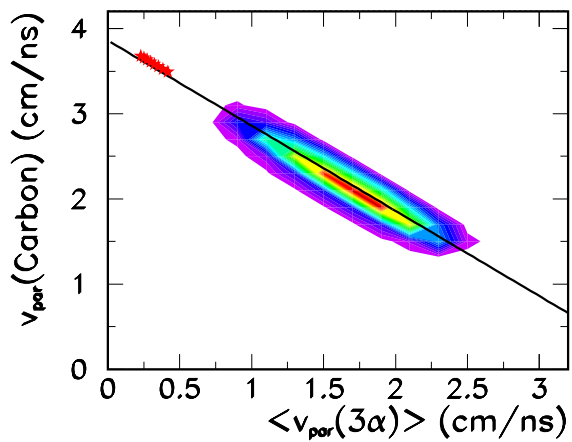}
\end{center}
\caption{(Color online) 
Logarithmic contour plot of the correlation between the laboratory
parallel velocity of the detected C residue and the laboratory
parallel velocity of the center of mass of 
the three $\alpha-$particles for all $Q_{kin}$ values. 
The solid line is the expected correlation from the parallel momentum
conservation in the laboratory system. The stars give the kinematical
locus of binary reactions compatible with the detection of a
quasi-projectile Carbon fragment in the experimental apparatus. 
}
\label{2c}
\end{figure}

This discrepancy in the $C$ channel may be due to a
preferential $\alpha-$particle emission by the reaction partners, followed by a
standard compound nucleus decay of the incomplete fused source. 
To investigate this hypothesis, we show in Figure~\ref{2c}
the correlation between the laboratory
parallel velocities of the detected $C$ residue and of the center of
mass of the three $\alpha-$particles. 
 
The correlation of Figure~\ref{2c} shows a single peak close to the
center of mass velocity ($\approx$ 2 cm/ns in this experiment). 
The kinematical locus associated with alpha emission 
in a pure dissipative two-body kinematics is indicated by the stars in
the figure.  
Contribution of peripheral binary reactions is therefore excluded. The
experimental distribution can rather suggest the decay from quasi-molecular
states of the excited $^{24}$Mg.  

Summarizing, only in the case of Oxygen a contamination from direct
reactions can contribute to the discrepancy between the data and the
statistical model. However, excluding the $Q_>$ less dissipative
events in the Oxygen channel, a large branching ratio discrepancy is
still observed.
This is shown by the dashed line in Figure~\ref{deltag}, where the
clusterization excess defined by (\ref{delta}) is evaluated only for
more dissipative events. 
Even in this case, an important branching ratio excess for
multiple $\alpha$ emission is observed with respect to the statistical
model. It is therefore clear that another mechanism is at play in
these channels and specific excited states with pronounced cluster
structures are populated.

In the next section we turn to study the kinematical
properties of the multiple $\alpha$ channels to better understand
the reaction mechanism or emission sequence leading to such events. 

\section{Multiple $\alpha$ correlations}\label{correlations}
\subsection{$^A$O-$\alpha$-$\alpha$ correlations}
\begin{figure}[h]
\begin{center}
\includegraphics[angle=0, width=0.65\columnwidth]{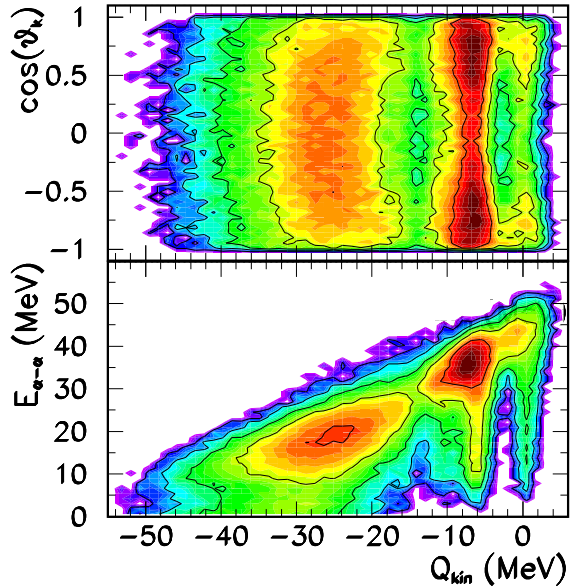}
\end{center}
\caption{(Color online) Logarithmic contour plot of the correlation
between $cos(\theta_k)$ (upper panel) and the $\alpha$-$\alpha$ relative
energy (lower panel) as a function of $Q_{kin}$ (\ref{qdef}) in
O-$\alpha$-$\alpha$ complete events.
} 
\label{jac1}
\end{figure}
To better understand the physical processes leading to the
$\alpha$-$\alpha$-$^A$O channels, we turn to study in greater detail
the topology of these three-body coincidence events as revealed by 
their kinematic correlations. 

Dealing with three-body systems, a useful representation can be obtained
making use of Jacobi energy-angular correlations, described in detail
in~\cite{demo}. 

We limit ourselves to the study of the correlations between the two
$\alpha-$particles in the “T”-system, where the core is the Oxygen.
Here, one of the coordinates is the cosinus of the relative angle
$\theta_k$ of the Oxygen residue momentum $\vec{k}_O$ 
and the $\alpha$-$\alpha$ relative momentum 
$\vec{k}_{\alpha-\alpha}$~\cite{ohlsen}:
$cos(\theta_k) = (\vec{k}_O \cdot
\vec{k}_{\alpha-\alpha})/(k_O  \ k_{\alpha-\alpha})$.\\
The other coordinate is $\epsilon = E_{\alpha-\alpha}/E_T$, i.e. the
ratio of the relative energy between the two $\alpha$'s to the total
available energy. 
In our case, for more dissipative $Q_{<}$ events the undetected neutron 
prevents the calculation of the normalized relative energy $\epsilon$, but
$cos(\theta_k)$ can still be used.

Figure~\ref{jac1} shows the $cos(\theta_k)$ (upper panel)
and the $\alpha$-$\alpha$ relative energy (lower panel) distributions 
as a function of $Q_{kin}$ (defined in (\ref{qdef})). 
\begin{figure}
\begin{center}
\includegraphics[angle=0, width=0.7\columnwidth]{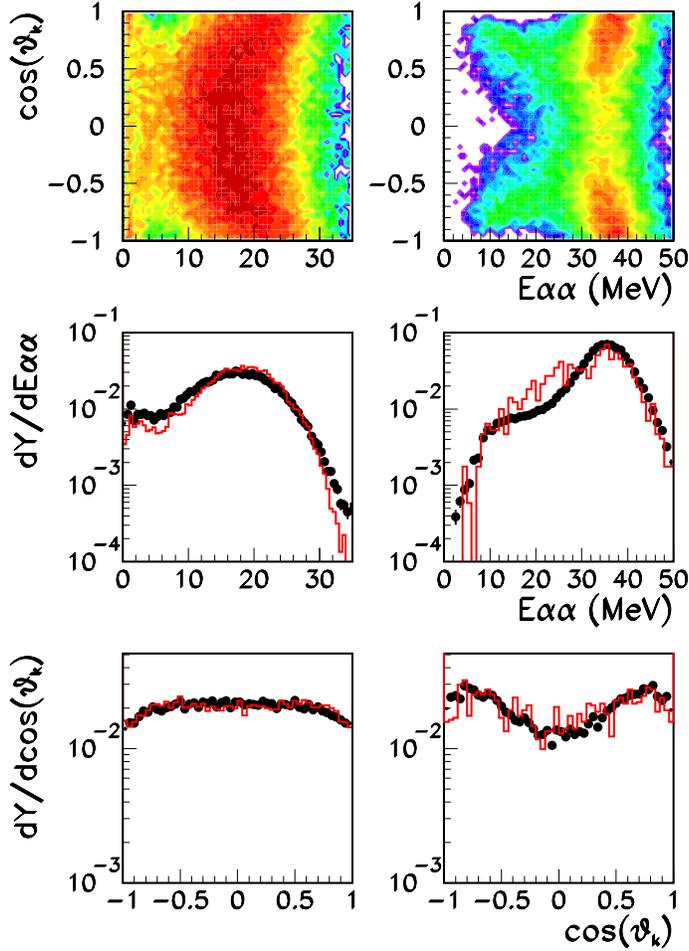}
\end{center}
\caption{(Color online) Left panels correspond to more dissipative and
right panels to less dissipative O-$\alpha$-$\alpha$ complete events.
Upper panels: Logarithmic contour plot of the correlation between
$cos(\theta_k)$ and the $\alpha$-$\alpha$ relative energy. 
Middle panels: projections of the $\alpha$-$\alpha$ relative
energy.
Lower panels: projections of the angular correlation $cos(\theta_k)$.
Data (symbols) are compared to HF$\ell$ calculations (lines).
All distributions are normalized to the unitary area. 
} \label{jac2}
\end{figure}

We can recognize that the two dissipation event classes $Q_<$ and
$Q_>$ defined in the previous section correspond to radically different
patterns of relative $\alpha$-$\alpha$ motion. Non-dissipative events are
characterized by a relative energy that can overcome $80\%$ of the
total available energy, and a back-to-back emission with respect to
the oxygen residue.  More dissipative events correspond
to particles closer in momentum space,
consistent with an increased memory loss of the entrance
channel. For these events, the $cos(\theta_k)$ angular distribution
is essentially flat.

The energy and angular correlations for the two classes of dissipation are
separately shown in Figure~\ref{jac2}, where data are compared to
model predictions.

As far as the relative $\alpha-\alpha$ energy is concerned (middle
panel) we can recognize two distinct contributions in
the $Q_>$ distribution. The dominant bump corresponds to the ground
state  of $^{16}O$, while the lower energy shoulder is due to the
population of the excited particle bound states, as we have already
observed commenting the $Q_{kin}$ distribution (Figure~\ref{jac1}). More
interesting, two contributions are also visible in the distribution
associated with more dissipative events. The small peak at low relative
energies suggests that a part of the observed 2-$\alpha$ 
emission might be associated with a correlated Be state. 
\begin{figure}[h]
\begin{center}
\includegraphics[angle=0,width=0.65\columnwidth]{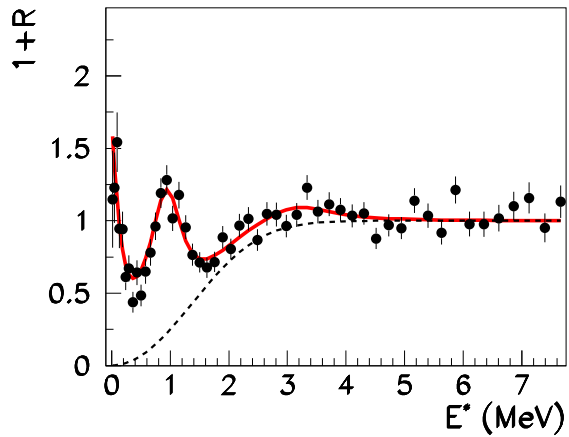}
\end{center}
\caption{(Color online) Correlation function of the relative $\alpha-\alpha$ 
energy in the $\alpha-\alpha-^A$O channel (symbols), and
corresponding fit (solid line) using the technique of~\cite{noi}. 
Dashed line: estimated Coulomb background.
}
\label{corraa}
\end{figure}

To explore this possibility, we analyzed the $\alpha-\alpha$ correlation 
function, obtained as the ratio of the measured and uncorrelated
distributions.
This correlation function, shown in Figure~\ref{corraa}, can be very
well fitted assuming a convolution of Breit-Wigner distributions
corresponding to the different low-lying $^8$Be and $^9$Be states. For
the details about the correlation function technique, see~\cite{noi}. 
A clear contribution of $^8$Be ground state and first excited
state (3.03 MeV) is visible in the figure, as well as the $^9$Be ground state 
at a relative energy of about 1 MeV.
The statistical model calculation produces a completely flat
correlation function and fails to reproduce these structures. 
It is important however to point out that events associated with the formation
of discrete $Be$ levels do not represent more than 3\% of
the experimental yield in the more dissipative $^A$O$-2\alpha$ channel, 
and can not be responsible for the global excess in the branching ratio.

Coming to the angular correlation of Figure~\ref{jac2}, in the case of
a simultaneous three-body event, it is expected that 
$cos(\theta_k)=\pm 1$ should be depleted due to the Coulomb
repulsion~\cite{demo}. 
Therefore, the observed flat distribution associated with $Q_<$
appears compatible with sequential emission. This is confirmed by the fact
that the shape of the distribution is well reproduced by the HF$\ell$
calculations. 
The sequential statistical model correctly reproduces also the
forward-backward peaked distribution of $Q_>$ events.  This shape can
be understood as due to angular momentum effects. Indeed a high
angular momentum of the emitting source  introduces a preferential
emission direction for the first evaporated particle. The relative
momentum vector $\vec{k}_{\alpha-\alpha}$ tends to keep this direction
if the second emission is isotropic, which happens if the first
particle takes away most of the initially available angular
momentum. 

\begin{figure}
\begin{center}
\includegraphics[angle=0, width=0.65\columnwidth]{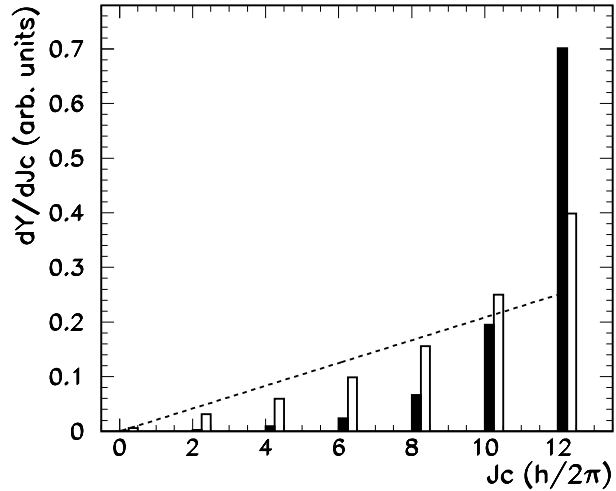}
\end{center}
\caption{(Color online) 
Compound angular momentum distribution in the unfiltered HF$\ell$
calculations associated with different emission channels containing
oxygen residues.  
More dissipative channels are represented by the white histogram, 
less dissipative by the black histogram.
A shift of 0.25 $\hbar$ has been applied to the $Q_<$ events for a
better representation. 
The distributions are normalized to the unitary area. 
The dashed line represents the global triangular distribution assumed
for the compound. 
} \label{jac2jc}
\end{figure}

This interpretation is confirmed by inspection of Figure~\ref{jac2jc}. 
This figure displays the compound angular momentum
distribution in the unfiltered HF$\ell$ calculations associated with
different emission channels containing Oxygen residues. We can see
that the angular momentum distribution is considerably steeper for
$^{16}$O$-\alpha-\alpha$, that is $Q_>$ events, with respect to the
global triangular distribution assumed for the compound. More
dissipative channels involving neutron emission are associated with a
higher thermal excitation energy, and therefore lower rotational
energy in order to conserve the total available compound nucleus
energy. The reduced angular momentum for this class of events can
therefore explain the observed isotropic $cos(\theta_k)$ distribution.
We can conclude that the Jacobi angular correlations of
both dissipation event classes are compatible with sequential
decay from an evaporation source.
However we recall that the branching ratio towards
two-$\alpha$ decay associated with the Oxygen residue shows an
important excess in the experiment compared to the HF$\ell$ calculation
for both classes of dissipation.
For this reason, we can not interpret the good reproduction of the
$cos(\theta_k)$ distribution by the model as a proof of standard
statistical decay in the data.

Concerning the $Q_>$ events, it is important to notice 
that the kinematical focussing is also compatible with the picture
of an incomplete fusion, with two $\alpha$ particles left over from
the two $\alpha$-clustered  collision partners. This latter mechanism is
indeed the most probable origin of  $(^{16}$O$-\alpha-\alpha$)
events, as we have argued in the previous section. This shows that the
interpretation of Jacobi kinematic correlations should be handled
with caution. 

Concerning the most dissipative $(^{A}$O$-\alpha-\alpha$) events,
the strong effect of angular momentum on the $cos(\theta_k)$ distribution
suggests that this observable can be used to explore the
effect of the undetected neutron(s) on the $\alpha-\alpha$ kinematical
correlations and gather further insight in the physical process at
play.

\begin{figure}[h]
\centering
\includegraphics[angle=0,width=0.9\columnwidth]{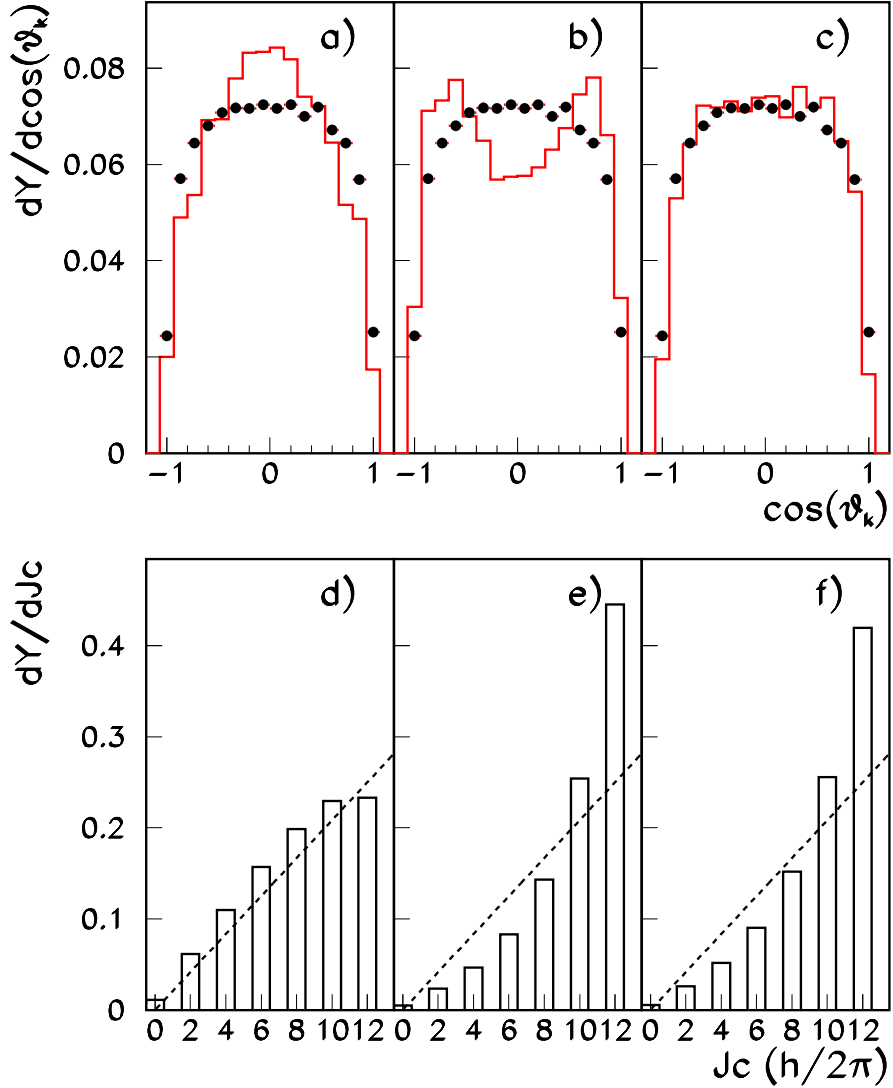}
\caption{(Color online) Analysis of dissipative $^{15}$O+2$\alpha$
complete events. 
Upper part: the experimental (dots) $cos(\theta_k)$ distribution
is compared to the calculated ones (red lines) according to different 
emission sequences (see text). Lower part: Compound angular momentum
distribution in the unfiltered HF$\ell$ calculations.
The dashed line represents the global triangular distribution assumed
for the compound. 
\\
Panel a) and d): n is the first emitted particle.
Panel b) and e): n is the second emitted particle.
Panel c) and f): n is the third emitted particle.
All distributions are normalized to unitary area.}
\label{costhetak-hf}
\centering
\end{figure}

This is shown in Figure~\ref{costhetak-hf}, where we compare the
experimental distribution with the HF$\ell$
predictions corresponding to different evaporation sequences.
The corresponding angular momentum distributions of the
initial $^{24}$Mg compound are also displayed in bottom panels of the
same figure.

When the neutron is the first particle emitted from $^{24}$Mg (panel a)
and the two $\alpha$'s correspond to the last two steps of the
de-excitation chain, the distribution is single-peaked. Indeed the
evaporation sequence corresponds to the lowest average angular
momentum for the initial compound (panel d), and moreover
the angular momentum available for the first $\alpha$
emission is reduced by the first step of the de-excitation.
Conversely, if the two $\alpha$'s are the first and last emitted particles,
with the neutron being emitted in the intermediate step (panel b),
the distribution shows two clear peaks.  
This is consistent with two sequential emissions from a high $J$ source
and a low $J$ source, as discussed above. An intermediate
situation is observed in panel c),
which displays the $cos(\theta_k)$ distribution obtained when two
successive $\alpha$'s are emitted from the excited $^{24}$Mg, and the
residual $^{16}$O is excited above the neutron separation energy,
leading to neutron emission in the last evaporation step. In this case
the initial angular distribution is compatible with the one of 
panel b), but  the angular momentum of the $^{20}$Ne is not
negligible. 
As a consequence, the second emission
leads to a modification of the direction of $\vec{k}_{\alpha-\alpha}$
with respect to the first $\alpha$ emission angle, 
and a smoothing of the two-peaked distribution.

From Figure~\ref{costhetak-hf} we can conclude that the measured
distributions are closer to the kinematical configuration expected for
the $\alpha-\alpha-$n sequence.
In turn, this implies  that
the preferential $\alpha$ emission observed in the data can be
attributed to the excited state of $^{24}$Mg and/or $^{20}$Ne. 

\subsection{$^A$C$-\alpha-\alpha-\alpha$ correlations}
In the case of the $(^AC$-3$\alpha$) channel, the presence of more
than three bodies in the exit channel in principle prevents from using
Jacobi observables.
\begin{figure}[h]
\centering
\includegraphics[angle=0, width=0.75\columnwidth]{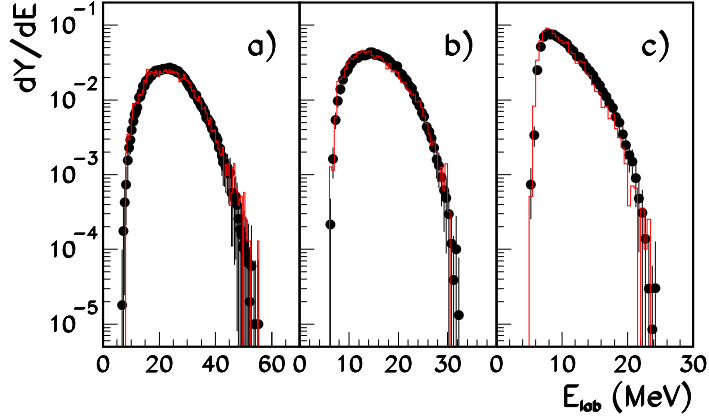}
\centering
\caption{(Color online) Energy spectra of $\alpha-$particles detected
in coincidence with a Carbon residue, ordered in each event
according to their laboratory energy as a fast (panel a), medium
(panel b) and slow (panel c).
Data (symbols) are compared to HF$\ell$ calculations (lines). 
All distributions are normalized to the unitary area. 
}
\label{e3a}
\end{figure}
 However, clear indications exist that a doorway $^8$Be
state is populated in the decay associated with this channel, leading
again to an effective three body $^{12}$C-$\alpha$-$^8$Be problem in
the absence of neutron emission, similarly to the Oxygen case. 
To demonstrate this statement, we have classified the three
coincident $\alpha-$particles in each event as slow, medium and fast
according to their laboratory energy~\cite{freer2,rana}. 

The energy distribution of $\alpha$-particles is very well
reproduced by the model, as shown in Figure~\ref{e3a}.
This indicates that the kinematics of the decay is well described by a
sequential evaporation mechanism. 

\begin{figure}[h]
\centering
\includegraphics[angle=0, width=1.1\columnwidth]{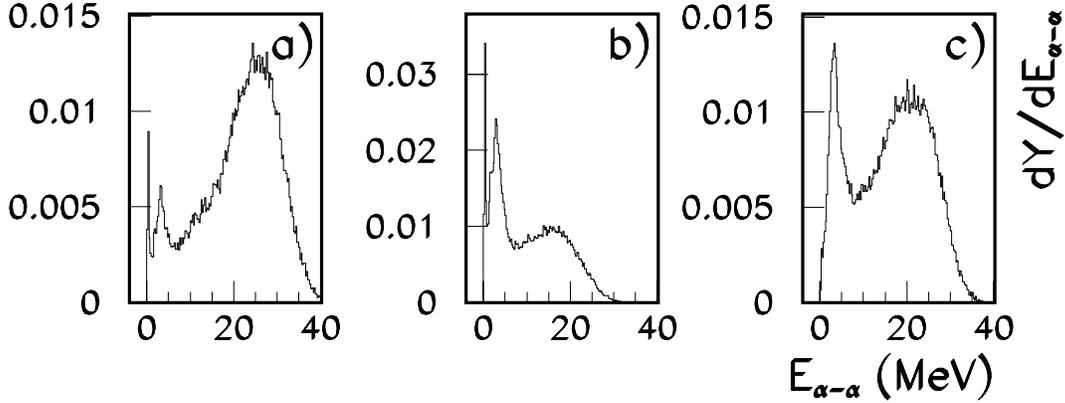}
\caption{Distribution of relative $\alpha$-$\alpha$ energy in the
($^A$C$-3\alpha$) channel. 
The three coincident $\alpha-$particles are ordered in each event
according to their laboratory energy as a slow, medium and fast
$\alpha$. Panel a) gives the experimental distribution of relative
energy between the fast and the medium, panel b) between the slow and
the medium, panel c) between the slow and the fast particle of each
event. All distributions are normalized to the unitary area.
} 
\label{erel_carbon}
\centering
\end{figure}

However, discrepancies between data and model appear in the
correlation observables.
Indeed, the experimental distributions of $\alpha-\alpha$ relative
energy of Figure~\ref{erel_carbon} show clear peaks corresponding to
the population of discrete excited states, the highest probability
being associated with the two slowest particles of each event. 
The number of events where two out of the three particles are found with a
relative energy below $6$ MeV is 40\% of the total sample. 
We can therefore safely argue that a doorway Be state is
frequently populated in these events.

\begin{figure}[h]
\centering
\includegraphics[angle=0, width=1.\columnwidth]{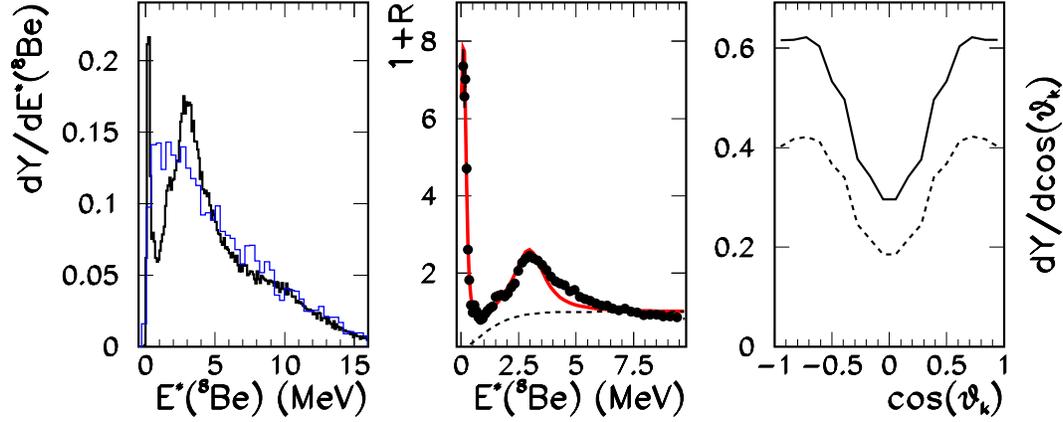}
\caption{(Color online) Left panel: Excitation energy distribution of
 the reconstructed $^8$Be$^*$ in the $^A$C$-3\alpha$ channel. In each
event, the two $\alpha-$particles with the smallest relative energy are
selected. The thin blue line represents HF$\ell$  predictions.
Middle panel: Correlation function of the $^8$Be$^*$ excitation energy
(symbols), and corresponding fit (solid line) using the technique
of~\cite{noi}. Dashed line: estimated Coulomb background.
Right panel: Jacobi $cos(\theta_k)$ distribution in the $T$-system
obtained replacing the the two $\alpha-$particles with the smallest relative
energy by their center-of-mass momentum. Full line: all
events. Dashed line: analysis limited to a reconstructed $^8$Be
excitation energy $E^*\leq 6$ MeV. 
All distributions are normalized to the number of events. 
} 
\label{jacobi_be}
\centering
\end{figure}

The excitation energy distribution of the $^8$Be$^*$, reconstructed
from the center-of-mass energy of the two $\alpha-$particles closer
in momentum space, is shown in the left part of Figure~\ref{jacobi_be}. 
The contribution of the ground and first excited $^8$Be state (3.03 MeV)
can be clearly recognized.
Model events, analyzed in the same way as the data and shown as a thin
blue line in the same panel, show a broad distribution without peaks
corresponding to excited states of $^8$Be.
As discussed in the previous paper~\cite{last_paper}, in the
Hauser-Feshbach formalism, light fragments have a negligible
probability to be emitted and are only obtained as evaporation residues. 
This discrepancy could be interpreted as the presence of a break-up
contribution in the data which is not properly treated by the
sequential calculation. 
Alternatively, it could indicate the existence of correlated alpha
structure in the excited even-even nuclei which are explored in the
de-excitation chain, as we have already suggested in
\S\ref{multiplealpha}. 

The contribution of these states is clearly visible in the correlation
function drawn in the middle panel, showing the fit through a
convolution of Breit-Wigner distributions corresponding to the
different low-lying $^8$Be states~\cite{noi}. 

The Jacobi $cos(\theta_k)$ distribution in the $T$-system, obtained
replacing the two $\alpha-$particles with the smallest relative energy by
their center-of-mass momentum, is displayed in the right part of
Figure~\ref{jacobi_be}. 
The observed kinematic focussing demonstrates that the process
is sequential. The existence of an uncorrelated background at $^8$Be
excitation energies higher than those expected from a population of
discrete levels does not change this conclusion. 
Indeed, the $cos(\theta_k)$ distribution is not affected by
limiting the analysis to $E^*\leq 6$ MeV for the $^8$Be spectrum
(Figure~\ref{jacobi_be} right panel). 

A complementary information on the dynamics of $\alpha$ emission
in the Carbon channel can be obtained by using the technique of Dalitz
plot~\cite{dalitz}. 
To do this, one has to build the Dalitz coordinates:
$x_D = \sqrt(3)/2 (Erel_{23} - Erel_{12})$ 
and
$y_D = (2 Erel_{13}- Erel_{23} - Erel_{12})/2$
where $Erel_{ij}$ is the relative energy of the i-th and j-th
particles.
\begin{figure}[h]
\centering
\includegraphics[angle=0, width=0.75\columnwidth]{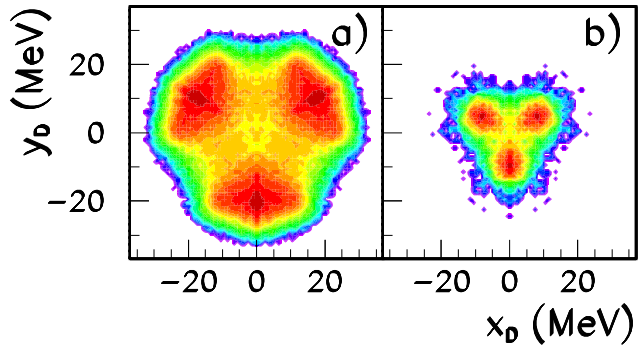}
\caption{(Color online) $\alpha$ Dalitz plot in the $^A$C$-3\alpha$ channel 
for low dissipation ($Q_>$) events (panel a) and high dissipation
($Q_<$) events (panel b).
 }
\label{dali}
\centering
\end{figure}
The Dalitz plots for the $3-\alpha$ coincidences are shown in
Figure~\ref{dali}, with different dissipation cuts. 
Independently of the dissipation, the Dalitz plot
clearly shows three bumps indicating a sequential process~\cite{dalitz}. 
The dissipation is obviously correlated with the excitation energy.
The excitation energy considered in the present case 
is much higher than the $3-\alpha$ Hoyle state of $^{12}$C, 
and discrete Carbon states decaying into three $\alpha-$particles
are not populated in the experimental sample, due to the high beam
energy and the selection of central events. 

The information given by the correlation studies reported in 
Figures~\ref{jacobi_be} and \ref{dali} consistently indicates 
a sequential evaporation of $\alpha-$particles from an
excited $^{24}$Mg leaving a Carbon residue, and tends to exclude the
excitation of a binary quasi-molecular ($^{12}$C+$^{12}$C$)^*$ state. 
Similarly to the Oxygen case, the main deviation from a standard
statistical behaviour concerns the branching ratio of the channel. 
This again indicates a preferential $\alpha$ emission from
the compound $^{24}$Mg and/or its daughter nucleus $^{20}$Ne. 
In the Carbon case, a non-negligible (40\%) fraction of the events
corresponds to a correlated alpha emission as a doorway $^8$Be. This
feature is also not reproduced by the statistical model. 

\subsection{6-$\alpha$ coincidences}\label{6a}
\begin{figure}[h]
\centering
\includegraphics[angle=0, width=0.9\columnwidth]{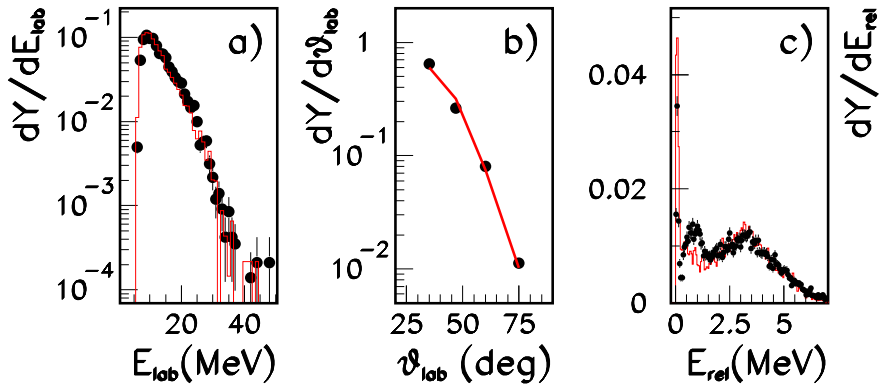}
\caption{(Color online) Analysis of $6-\alpha$ coincidences (symbols),
compared to HF$\ell$ predictions (lines). Panel a): $\alpha$ energy
spectrum. Panel b): $\alpha$ angular distribution. Panel c):
 $\alpha-\alpha$ relative energy distribution.
All distributions are normalized to unitary area. 
}
\label{6alpha}
\centering
\end{figure}
Finally, we turn to examine events where the whole available mass and charge is 
found as $\alpha-$particles. The branching ratio of this channel
is well reproduced by the statistical model, as we have already
observed in Figure~\ref{multa}. It is however interesting to check if the
kinematical properties of the observed events are compatible with a
sequential $\alpha$ emission. 

The ensemble of the different $\alpha$ observables are presented in
Figure~\ref{6alpha}, in comparison with the statistical model
predictions. We can see that the model successfully reproduces the
kinematic characteristics of these events. 
In panel c) we can recognize the ground state and the first excited
state of $^8$Be (3.03 MeV), which are well reproduced by the model. 
The experimental distribution of the $\alpha-\alpha$ relative energy
shows an extra-peak at about 1 MeV, not seen in the model.
It corresponds to the ground state of $^9$Be and it is due to a
$^{3}$He$-\alpha$ contamination of $Z=2$ particles at the limit of
the mass identification threshold, as already observed in
Figure~\ref{q_exp_12}. The contribution of this contamination has been
evaluated to be about 7\%.
If we ignore this spurious contribution, we can estimate that the
events showing a $^8$Be correlation with an excitation energy below $6$ MeV
are 90\% of the $6-\alpha-$particles sample, in agreement with the
statistical model.  
The presence of these Beryllium correlations shows that the
six-$\alpha$ emission is fully compatible with a sequence of binary
decays, with the last evaporation step leaving an unstable $^8$Be
residue~\cite{dipietro}. 
%
%

\section{Conclusions}\label{conclusions}
The $^{12}$C$(^{12}$C$,X)$ reaction has been studied at 95 MeV beam
energy with the GARFIELD+RCo experimental setup at LNL-INFN. 
Events completely detected in charge ($Z_{tot}=12$) have been selected
and compared to Hauser Feshbach statistical model calculations
performed with the code HF$\ell$.

Some clear deviations from a statistical behaviour in the decay 
have been found. These deviations concern an anomalously high
branching ratio towards multiple (two or three) $\alpha$ emission.
Conversely, the probability of a complete vaporisation into six
$\alpha-$particles is well reproduced. This latter channel is shown to
be compatible with a standard statistical sequential emission, leading
to a final unstable $^8$Be evaporation residue. 

Two different phenomena can explain the extra yield
in the 2 and 3 $\alpha-$particle exit channels.
First, an extra experimental cross section for the
three-body less dissipative $^{16}$O$-\alpha-\alpha$ decay channel 
has been attributed to the contamination of direct reactions.
This reaction mechanism mixing is not the unique source of discrepancy
with the HF$\ell$ predictions. 

Despite these events have been excluded and
all the kinematic characteristics were reproduced, the branching
ratios of the multiple $\alpha-$decay channels were found to be still
largely underestimated by the calculation. 

A detailed analysis of the multiple $\alpha-$particle correlations in
these channels indicates a sequential process with a clear hierarchy
in the emission sequence. 
The highest probability is associated with the first chance
emission of an $\alpha-$particle from a highly excited $^{24}$Mg
compound, leaving the daughter $^{20}$Ne nucleus still well above its
particle emission threshold. 
Then, the excited $^{20}$Ne preferentially emits another $\alpha-$particle,
leaving an Oxygen final evaporation residue, or alternatively an
excited $^8$Be nucleus in coincidence with a Carbon residue.
Neutron emission was not directly measured, but it was attributed to
the last emission step, according to kinematical correlations.
As a general conclusion, the persistence of cluster structures for
$^{24}$Mg and/or its daughter nucleus $^{20}$Ne, at excitation
energies well above the energy threshold for full disintegration into
$\alpha-$particles, can be inferred. 


\textit{Acknowledgments}\\
The authors thank the crew of the XTU TANDEM acceleration system at LNL. 
\\
This work was partially supported by the European Funds for Large
Scale Facilities - Seventh Framework Program - ENSAR 262010 and by
grants of Italian Ministry of Education, University and Research under
contract PRIN 2010-2011.

\end{document}